\documentclass[strucabstract]{aa}

\usepackage[]{natbib,amsmath,amssymb}
\bibpunct{(}{)}{;}{a}{}{,}
\usepackage{graphicx}
\usepackage{epsfig}

\newcommand{\bea}{\begin{eqnarray}}
\newcommand{\eea}{\end{eqnarray}}
\newcommand{\be}{\begin{equation}}
\newcommand{\ee}{\end{equation}}

\newcommand{\vc}[1]{\mbox{\boldmath $#1$}}
\renewcommand{\d}{{\mathrm d}}

\newcommand{\abs}[1]{\left| #1 \right|}
\newcommand{\rund}[1]{\left(#1\right)}
\newcommand{\eck}[1]{\left[ #1 \right]}
\newcommand{\ave}[1]{\left\langle #1 \right\rangle}

\def\elabel#1{\label{eq:#1}}
\begin{document}
  \title{Mass reconstruction by gravitational shear and flexion}

  \author{Xinzhong Er \inst{1,2}
    \and Guoliang Li \inst{1,3}
    \and Peter Schneider \inst{1}
  }

  \institute{ Argelander-Institut f\"ur Astronomie, Universit\"at Bonn,
    Auf dem H\"ugel 71, D-53121 Bonn, Germany
    \email{xer@astro.uni-bonn.de}
    \and
    International Max Planck Research School (IMPRS)
    for Astronomy and Astrophysics
    Auf dem H\"ugel 69, D-53121 Bonn, Germany
    \and
    Department of Physics, Purdue University, 525 Northwestern Avenue,
    West Lafayette, IN 47907, USA
  }

  \date{Received --; accepted-- }

  \abstract{}
       {Galaxy clusters are considered as excellent probes for
             cosmology. For that purpose, their mass
  needs to be measured and their structural properties needs to be
  understood.}
       {We propose a method for galaxy cluster mass reconstruction which
             combines information from strong lensing, weak lensing shear and
             flexion. We extend the weak lensing analysis to the inner parts
         of the cluster and, in particular,
         improve the resolution of substructure.}
       {We use simulations to show that the method recovers the mass
         density profiles of the cluster. We find that the weak lensing
         flexion is sensitive to substructure. After combining the
         flexion data into the joint weak and strong lensing analysis,
         we can resolve
         the cluster properties with substructures.
       }
       {}

  \titlerunning{Shear and Flexion United}
  \authorrunning{X.Er et al.}

  \maketitle

\section{Introduction}

Galaxy clusters are the most massive virialized structures in the
universe, and are considered as valuable probes for cosmological
parameters.  From numerical simulations one can identify collapsed
structures (called halos) and study their properties. In $\Lambda$CDM
simulation, the density profile of halos has a universal form, closely
following the NFW profile \citep{1996ApJ...462..563N}. Moreover,
numerical simulations predict a rich population of sub-clumps within the
virial radius of the main halo.

Weak gravitational lensing provides a powerful tool for studying the
mass distribution of clusters of galaxies as well as the large-scale
structure in the Universe
\citep[see][for reviews on weak lensing]{2001PhR...340..291B,
  2003ARA&A..41..645R, 2006glsw.book.....S, 2008PhR...462...67M}.
Several methods were developed on cluster mass reconstructions
\citep[see e.g.][]{1993ApJ...404..441K,2001A&A...374..740S}.
Besides weak lensing, strong lensing systems are also employed to extend
the weak lensing analysis into the inner part of clusters, such as the
multiply-imaged systems and the corresponding identification of the
location of critical curves
\citep{2006A&A...458..349C, 2005A&A...437...39B}.
Since then some cluster mass reconstructions have been successfully carried out
\citep{2005A&A...437...49B,2006ApJ...652..937B,2009A&A...500..681M}.
Both methods reconstruct the gravitational potential $\psi$, based on
minimizing a $\chi^2-$function comparing observed data with model
\citep{1996A&A...313..697B,1998A&A...337..325S}.

Flexion has been recently studied as the derivative of the shear, and
responds to small-scale variations in the gravitational potential
\citep{2002ApJ...564...65G,2005ApJ...619..741G,2006MNRAS.365..414B}.
The two complex spin-1 and spin-3 components of flexion together with
the shear describe how intrinsically round sources are mapped into
`arclets'.  The measurement of flexion can be obtained in principle
from the same imaging data as used for shear measurements, and methods
for that have been discussed in the literature
\citep{2006ApJ...645...17I,2007ApJ...660..995O,
  2007ApJ...660.1003G,2008A&A...485..363S}.  One disadvantage of shear
technique is that galaxies have a broad intrinsic ellipticity
distribution which constitutes a serious noise component.
Since third-order brightness moments of galaxies are suspected to be
much smaller than second-order ones, we expect a lower intrinsic noise
in flexion measurements.
It has been noted that flexion can contribute to study the galaxy and cluster
halo mass and density profiles, and particularly sensitive to substructure
\citep{2009arXiv0909.5133B,2009MNRAS.395.1438L,2008ApJ...680....1O}.
\citet{2008ApJ...680....1O,2007ApJ...666...51L} have used flexion
measurements to detect substructure in Abell 1689.

We propose a method that combines shear measurements and flexion into
a weak lensing mass reconstruction, which enable us to reconstruct the
mass distribution of a lens with special sensitivity to
substructure. It is based on an estimations of the lensing potential
using $\chi^2$-minimization. The resulting equations are first
linearzied and then solved by iteration.  We outline the basic lensing
formalism in Sect.\ts\ref{sc:2}, and the ideas of reconstruction
method in Sect.\ts\ref{sc:method} (calculations are shown in the
Appendix). Numerical tests are presented in Sect.\ts\ref{sc:test}, and
conclusions are given in Sect.\ts\ref{sc:conclusion}.

\section{\label{sc:2}Basic formalism}

We mainly follow the notations of \citet{2008A&A...485..363S} to
present the basic gravitational lensing shear and flexion formalism.
We adopt the thin lens approximation where the lensing mass
distribution is projected onto the lens plane.  Thus, all the lensing
properties are contained in the lensing potential $\psi$, which is the
projected Newtonian gravitational potential on the lens plane
\be
\psi (\vc\theta ) =
{1 \over \pi} \int_{\Re^2} \d^2\theta' \kappa(\vc\theta ')
{\rm ln}|\vc\theta-\vc\theta'| \;,
\ee
where $\vc\theta$ denotes the (angular) position in the lens plane,
and $\kappa(\vc\theta)$ is the projected mass density
$\Sigma(\vc\theta)$ of the lens in units of
\be
\Sigma_{\rm cr}={c^2 \over 4 \pi G}{D_{\infty} \over D_{\rm d}D_{\rm d,\infty}},
\;\;{\rm so}\;\;\kappa(\vc\theta) = {\Sigma(\vc\theta) \over \Sigma_{\rm cr}}
\ee
for a fiducial source located at a redshift $z\rightarrow \infty$.
Here $D_{\infty},D_{\rm d}$ and $D_{\rm d,\infty}$ are the angular
diameter distances between the observer and the source, the observer
and the lens and between the lens and the source, respectively.  The
second-order local expansion of the lensing equation in Cartesian
coordinates reads
$\beta_i=\theta_i-\psi_{,ij}\theta_j-\psi_{,ijk}\theta_j\theta_k/2$,
where indices separated by a comma denote partial derivatives with
respect to $\theta_i$, and the Einstein summation convention is used.
The surface mass density $\kappa$ and the complex shear $\gamma
=\gamma_1 + {\rm i}\gamma_2$ are given in terms of the deflection
potential through
\be
\kappa = (\psi_{,11} + \psi_{,22})/2,\;
\gamma = (\psi_{,11} - \psi_{,22})/2 +{\rm i}\psi_{,12}\;.
\ee
The two flexions ${\cal F}, {\cal G}$ are combinations of third-order
derivatives of $\psi$, and also related to the gradient of $\kappa$
and $\gamma$
\bea
{\cal F}&=&{1\over 2}\eck{\psi_{,111}+\psi_{,122}+{\rm
i}\rund{\psi_{,112}+\psi_{,222}}} \; ; \nonumber\\
{\cal G}&=&{1\over 2}\eck{\psi_{,111}-3\psi_{,122}+{\rm
i}\rund{3\psi_{,112}-\psi_{,222}}} \; ,\nonumber\\
{\cal F}&=& \nabla_{\rm c} \kappa = \nabla_{\rm c}^* \gamma,\;
{\cal G} = \nabla_{\rm c} \gamma,
\elabel{kapdiff}
\eea
where we defined the differential operators
\be
\nabla_{\rm c} = {\partial \over \partial \theta_1} + {\rm i}
{\partial \over \partial \theta_2}; \;
\nabla_{\rm c}^* = {\partial \over \partial \theta_1} - {\rm i}
{\partial \over \partial \theta_2}.
\nonumber
\ee
The second-order lensing equation in complex notation then reads
\be
\beta=(1-\kappa)\theta-\gamma\theta^*
- {1\over 4}{\cal F}^*\,\theta^2
-{1\over 2}{\cal F}\,\theta\theta^*
- {1\over 4}{\cal G}\,(\theta^*)^2 \;.
\elabel{lenseq}
\ee
Due to the mass-sheet degeneracy in lensing system \citep[see, e.g.,][]
{1985ApJ...289L...1F,1995A&A...294..411S}
the lens equation is rewritten in reduced form as (see Schneider \& Er
2008)
\bea
\hat\beta &\equiv& {\beta\over (1-\kappa)} \\
&=& \theta-g\theta^* - \Psi_1^*(G_1)\,\theta^2
-2\Psi_1(G_1)  \,\theta\theta^*
- \Psi_3(G_1,G_3) \,(\theta^*)^2\;, \nonumber
\eea
where $g,G_{1},G_{3}$ are the reduced shear and the reduced flexion
components,
\bea
g&=&{\gamma \over 1- \kappa}\;;\nonumber \\
G_1&\equiv& \nabla_{\rm c}^* g={{\cal F}+g{\cal F}^*\over (1-\kappa)}\;;
\nonumber \\
G_3&\equiv& \nabla_{\rm c} g={{\cal G}+g{\cal F}\over (1-\kappa)}\;,
\elabel{G13def}
\eea
and where $\Psi_1,\Psi_3$ are given by
\bea
\Psi_1&=& {G_1-gG_1^* \over 4(1-gg^*)},\nonumber\\
\Psi_3&=& {G_3\over 4} - {g(G_1 - gG_1^*) \over 4(1-gg^*)}\;.
\eea
All the relation above are written for sources located at redshift
$z\rightarrow \infty$. For a source at redshift $z$ and lens at redshift
$z_{\rm d}$, the lensing strength is reduced by multiplying $\psi$
(and all its derivatives) by a \lq cosmological weight' function,
\be
Z(z)\equiv \eck{D_{\rm d,\infty} \over D_{\infty}}^{-1}{ D_{\rm ds} \over
  D_{\rm s}} H(z-z_{\rm d}),
\ee
where $D_{\rm ds}$ and $D_{\rm s}$ are the angular diameter distances between
the lens and the source, and the observer and the source.
$H(z-z_{\rm d})$ is the Heaviside step function to give zero weight to
sources located in front of the lens. Thus, at redshift $z$
\bea
\kappa(z) &=& Z(z) \kappa, \, \gamma(z) = Z(z) \gamma, \nonumber\\
{\cal F}(z) &=& Z(z) {\cal F}, \, {\cal G}(z) = Z(z) {\cal G}.
\eea

Information about the lensing potential can be obtained through the
reduced shear and reduced flexion by measuring the ellipticity
$\epsilon$ and the higher-order brightness moments, or called HOLICs
\citep{2007ApJ...660..995O}.  The information on the lens potential is
contained in the transformation between the source and image
ellipticity
\be
\epsilon^{\rm s} =
\begin{cases}
  \dfrac{\epsilon - g}{1- g^* \epsilon} \quad {\rm for}\;
  \bigl\lvert g \bigr\rvert <1;\\[1.5em]
  \dfrac{1- g \epsilon^*}{\epsilon^* -g^*} \quad {\rm for}\;
  \bigl\lvert g \bigr\rvert >1,\\
\end{cases}
\ee
and the higher-order brightness moments
\bea
T_1^{\rm s} &\approx&  T_1-{9 F_0-12 Q_0^2 \over 4}G_1,\nonumber\\
T_3^{\rm s} &\approx&  T_3- {3 F_0 \over 4} G_3,
\eea
where $Q,T,F$ are the second-, third- and fourth- order brightness
moments, respectively \citep[see][for
definitions]{2008A&A...485..363S}, the superscript $^{\rm s}$
indicates source quantities, and the subscript indicates the spin of
the moments. If we assume that the ensemble average of intrinsic
moments vanish, $\langle\epsilon^s\rangle=0$, $\langle T_{1,3}^{\rm
  s}\rangle=0$, the expectation value for the galaxy image quantities
yields the estimator of reduced shear and reduced flexion
\be
g =
\begin{cases}
  \bigl\langle \epsilon \bigr\rangle  \quad {\rm if } \;|g|<1, \\[1em]
  \bigl\langle {1 / \epsilon^*} \bigr\rangle, \quad {\rm otherwise},\\
\end{cases}
\ee
\bea
G_1 &\approx& t_1 = \ave{ {4 \over 9F_0-12Q_0^2}T_1};\nonumber\\
G_3 &\approx& t_3 = \ave{ {4\over 3 F_0}T_3 }.
\elabel{eg13}
\eea
Here we use $t_{1,3}$ to indicate the estimators of the reduced
flexion.  These estimates of reduced shear and reduced flexion break
down when the image is located on or close to a critical curve.

\section{\label{sc:method}Mass reconstruction method}

The strategy of our method is similar to that of the strong- and
weak-lensing united method \citep{2005A&A...437...39B,2005A&A...437...49B}.
Thus, we include flexion in the reconstruction of
the lensing potential $\psi$.  There are two reasons for combining
shear and flexion other than using flexion alone in the mass
reconstruction. In current observations, the number density of images
from which flexion can be estimated is low, as we can see in our
simulation data. Even with future higher flexion number density
surveys, flexion is most sensitive to substructures, but rather
insensitive to the larger-scale (cluster-scale) smooth potential.

\subsection{\label{sc:3.1}The $\chi^2$-function}

We describe the cluster mass distribution by
the deflection potential $\psi$ on a regular grid,
use finite differencing to calculate the deflection angle,
the reduced shear and the flexion. These quantities on
the grid are considered to describe the lens model which can be
compared with data. Our aim is to
seek a potential ${\psi}$ that minimizes the difference between model
($g,G_1,G_3$) and data ($\epsilon,t_1,t_3$).
We therefore define a $\chi^2$-function
\be
\chi^2(\psi)=\chi^2_s(\psi)+\chi^2_{\epsilon}(\psi)+\chi^2_{f}(\psi)
+ \eta R(\psi),
\ee
where $\chi^2_s$,
$\chi^2_{\epsilon}(\psi)$ and $\chi^2_{f}(\psi)$ contain the information from
strong lensing multiple image systems, shear and flexion, respectively.
$\eta R$ is a regularization term, used to
smooth out small-scale noise components, i.e., to avoid overfitting
the data.
For a multiple image system with $N_m$ images located at $\vc\theta_m$,
the $\chi^2_s$ term is well defined by
\citet{2005A&A...437...39B}
\be
\chi^2_s=\sum_{\rm str} \sum_{m=1}^{N_m} b_m^TS^{-1}b_m,
\ee
where $\sum_{\rm str}$ is the sum over all strong lens system, and
$\vc b_m =\vc \theta_m - \vc\alpha(\vc \theta_m) -\vc\beta_s$.
$\vc\beta_s$ is the average source position, and $S$ is the covariance matrix
$S={\rm diag}(\sigma^2_{s1}, \sigma^2_{s2})$. The shear term
$\chi^2_{\epsilon}$ is
\be
\chi^2_{\epsilon}=\sum^{N_{\rm g}}_{i=1}{|\epsilon_i - g(\vc \theta, z_i)|^2
  \over \sigma_{\epsilon}^2},
\elabel{epschi}
\ee
where $N_{\rm g}$ is the number of background galaxies, and
\be
\sigma_{\epsilon}^2=\rund{1-|g|^2}^2 \sigma_{\epsilon^s}^2+\sigma_{\rm err}^2,
\elabel{esigma}
\ee
with $\sigma_{\epsilon^s}\approx 0.3$ is the standard deviation
\citep{1996ApJ...466..623B}, $\sigma_{\rm err}$ is the measurement
error, which we take 0.1 in this paper, and $g$ is the reconstructed
reduced shear.

The flexion term is defined in a way similar to the shear
\be
\chi^2_f=\sum^{N_f}_{i=1}\left({|t_{1i}-G_{1}(\vc\theta_i,z_i)|^2 \over
  {\sigma_{t1}^2}} + {|t_{3i}-G_{3}(\vc\theta_i,z_i)|^2 \over
  \sigma_{\large t3}^2}\right),
\elabel{tchi}
\ee
where $G_{1,3}$ are the reduced flexions at position
$\vc\theta_i$ and redshift $z_i$ in our model.
Note that we can measure only the reduced flexion.
There is a significant difference between the reduced flexion and its
approximation ${\cal F}/(1-\kappa)$ or ${\cal G}/(1-\kappa)$ in the region
where the shear is not small.
However, the definition of reduced flexion
(Eq.\ref{eq:G13def}) renders the $\chi^2_f$ function (Eq.\ref{eq:tchi})
complicated and the equations become difficult to solve. We thus define
\be
G_1' \equiv {{\cal F} \over 1-\kappa}; \;\;\;
G_3' \equiv {{\cal G} \over 1-\kappa}.
\ee
From Eq.(\ref{eq:G13def}) it is easy to obtain
\be
G_1' = {G_1 - gG_1^* \over 1-gg^*}; \;\;\;
G_3' = G_3 - gG_1'.
\ee
We thus use a modified estimator for the observed reduced flexion,
\be
t_1'={t_1-\epsilon t_1^* \over 1-\epsilon\epsilon^*}; \;\;\;
t_3'= t_3-\epsilon t_1',
\elabel{reducedt1}
\ee
and replace $t_1$, $t_3$ by $t_1'$, $t_3'$ in (Eq.\ref{eq:tchi}). The flexion
term $\chi^2_f$ is thus redefined as
\be
\chi^2_f=\sum^{N_f}_{i=1}\left(\dfrac{\abs{\dfrac{{\cal F}_{i}}{1-\kappa_i}
  -t'_{1i}}^2}{\sigma_{t1'}^2} + \dfrac{\abs{\dfrac{{\cal G}_{i}}
  {1-\kappa_i}-t'_{3i}}^2}{\sigma_{t3'}^2}\right).
\elabel{tchi2}
\ee
The $\sigma^2_{t1'}$ and $\sigma^2_{t3'}$ are different from the
dispersion in flexion measurement or intrinsic flexion variance, which are
both difficult to obtain from current observations.
In \citet{2007ApJ...660.1003G}, the intrinsic scatter of
flexion are estimated as $\sigma_{a|{\cal F}|}=0.03$ and
$\sigma_{a|{\cal G}|}=0.04$,
where $a$ is the semi-major axis of the lensed image, so that the
combination $a|\cal F|$ represents a dimensionless term.
\citet{2009MNRAS.400.1132H,2009MNRAS.395.1438L} used a conservative estimate
$\sigma_{\cal F}=0.1/''$. Here we use
\bea
\sigma^2_{t1'}= \abs{{\partial G_1'\over \partial g}}^2 \sigma^2_{\epsilon} +
\sigma^2_{t1}; \nonumber\\
\sigma^2_{t3'}= \abs{{\partial G_3'\over \partial g}}^2 \sigma^2_{\epsilon} +
\sigma^2_{t3},
\elabel{fsigma}
\eea
where $\sigma_{t1}$, $\sigma_{t3}$ are the dispersion of flexion data.
The first term on the right is scatter due to intrinsic ellipticity and shear
measurement error.

In principle, the total number of flexion measurements is the same as
the number of galaxy images $N_f=N_g$ from which shear can be
estimated. But in reality, flexion is measurable only for a subset
of (larger) images; beside that, we discard low signal-to-noise
flexion measurements.  Thus the number of images from which flexion
can be estimated is usually smaller than the number of galaxy images
with shear measurements.

To find the minimum $\chi^2$-function, we solve the set of equations
\be
{\partial\chi^2(\{\psi_i\}) \over \partial \psi_i}=
0,
\elabel{chi2}
\ee
which is in general a non-linear set of equations.  This problem is
solved by an iterative procedure \citep{2005A&A...437...39B}.  We
perform a three-level iteration process, start with initial model
$\kappa^0$ as the fixed regularization, linearize the system
(Appendix), solve the linear system of equations as the
inner-level. These steps are repeated until convergence of $\kappa$ is
achieved. The middle-level is repeating the inner level with new
regularization, based on the estimate of $\kappa$ from previous step
result, until we can get $\chi^2_{\rm red} \sim 1$.  Finally we
increase the number of grid points in the field and repeat the first
two level iterations on the new grid until we reach the final desired
grid spacing, i.e., resolution.

For the regularization we choose
\be
R=\sum_{i,j=1}^{N_x,N_y} \eck{\kappa_{ij}^{(n)} -
  \kappa^{(n-1)}(\vc\theta_{ij})}^2,
\ee
where $\kappa^{(n)}$ and $\kappa^{(n-1)}$ are current and previous resulting
$\kappa$ map in every middle-level. For the first step, we can use an initial
model $\kappa^{(0)}$ which is obtained from other methods, or simple set
$\kappa^{(0)}=0$ across the whole field.


\subsection{Grid point field}

\begin{figure}
\centerline{\scalebox{0.5}
  {\includegraphics[width=7.5cm,height=7.5cm]{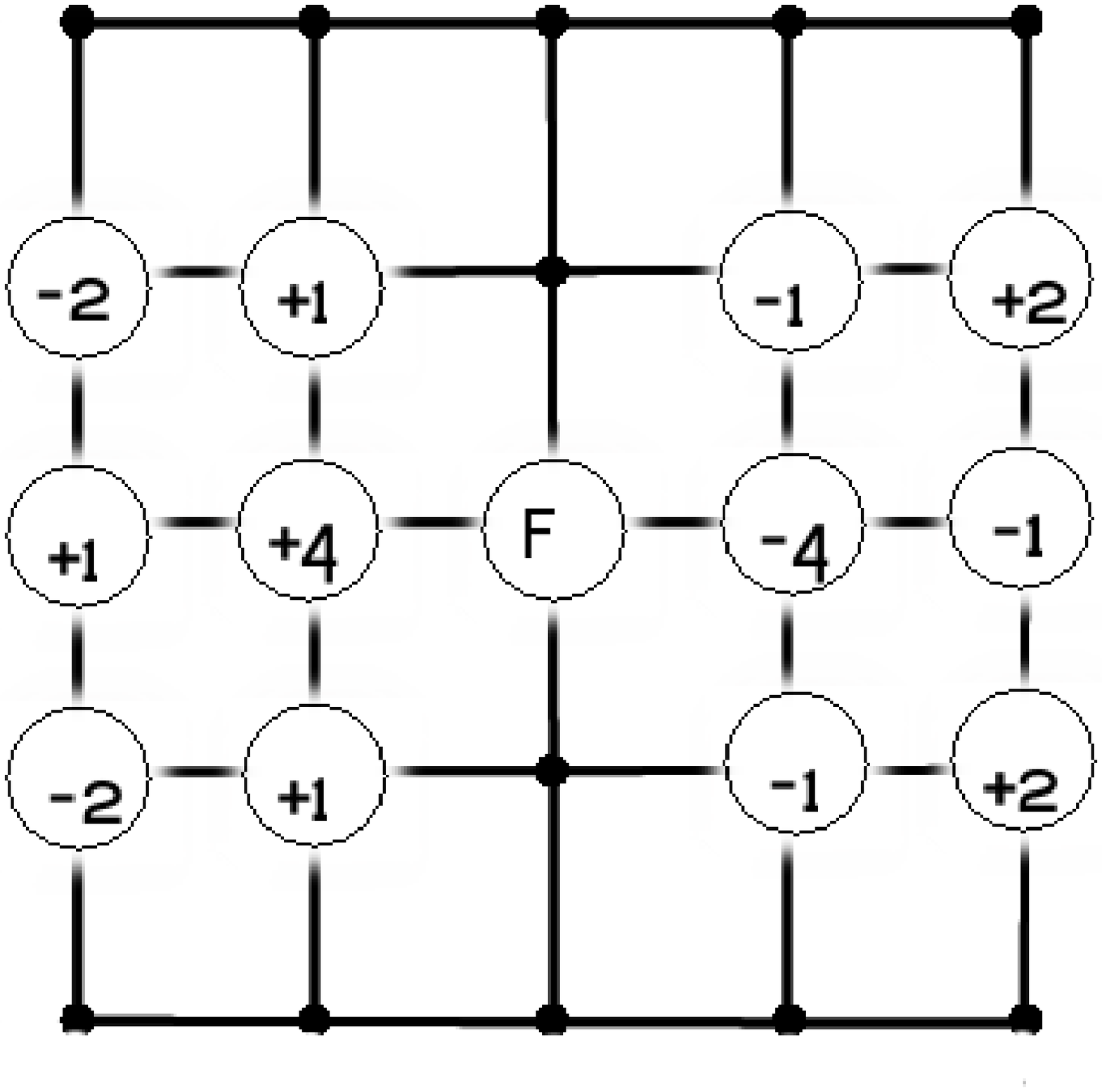}
    \includegraphics[width=7.5cm,height=7.5cm]{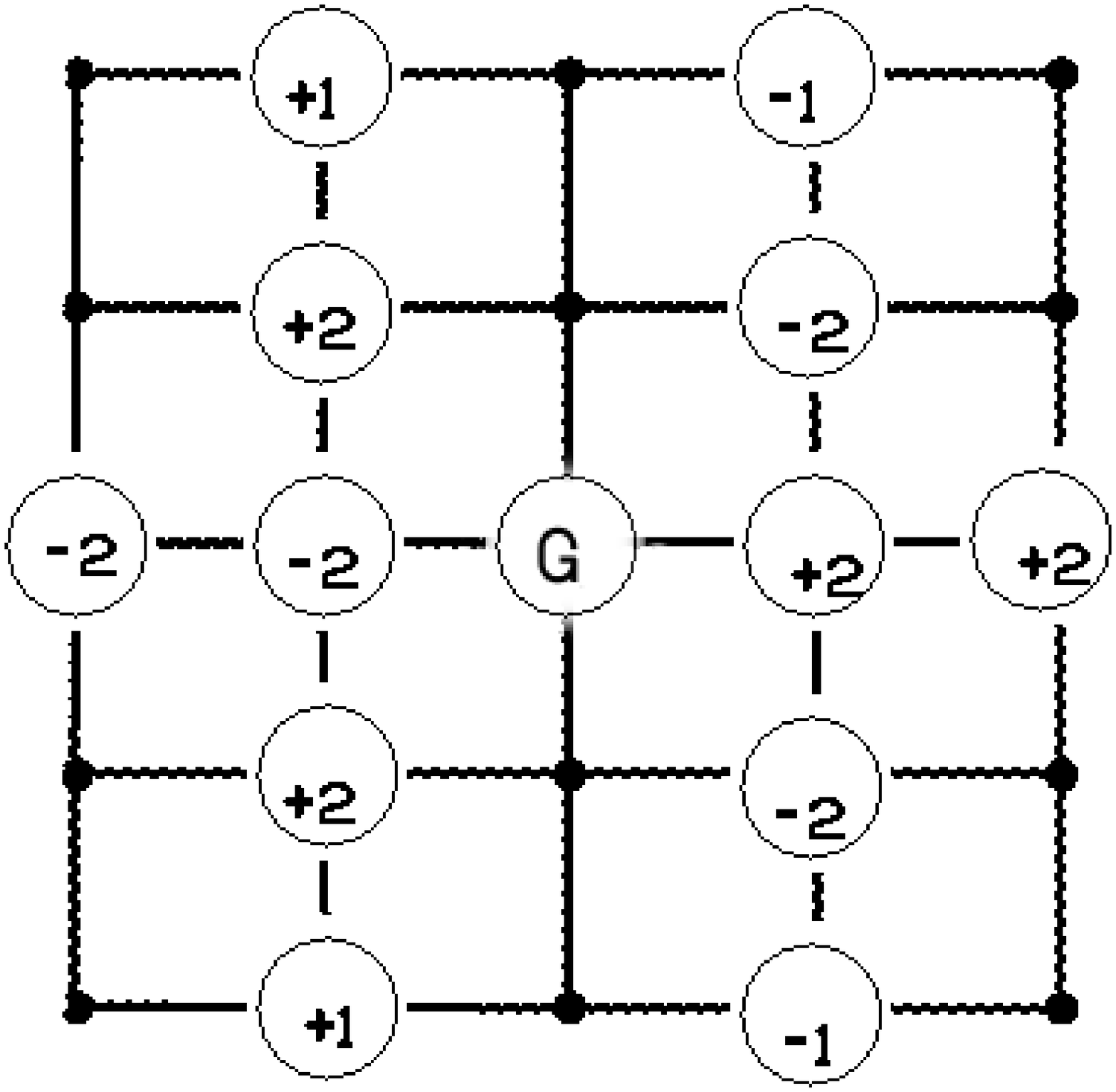}
}}
\begin{minipage}{4.3cm}
\centerline{${\cal F}_1 \to 1/(12 \Delta^3)$}
\end{minipage}
\begin{minipage}{4.3cm}
\centerline{${\cal G}_1 \to 1/(8 \Delta^3)$}
\end{minipage}
\caption{
The finite differencing coefficients of ${\cal F}_{1}$(left) and ${\cal
  G}_{1}$(right). The coefficients of ${\cal F}_2$(${\cal G}_2$) are the
same as that of ${\cal F}_1$(${\cal G}_1$) after rotating by $\pi/2$
anticlockwise(clockwise).
}
\label{fig:grid}
\end{figure}

Finite differencing techniques \citep{1972hmfw.book.....A} provide a
way to calculate $\kappa$ and $\gamma$ on the potential $\psi$ grid
field.  We need 9 grid points for second-order derivatives required
for $\kappa$ and $\gamma$, and 16 points for flexion. But by using
$4\times 4$ points, one would obtain the flexion value at the center
of a grid cell. Thus we instead use $5\times 5$ finite difference
scheme to obtain the flexion on the grid. For instance,
\be
{\cal F}_1(i,j)={1 \over 12\Delta^3} \sum^{+2}_{\Delta_i,\Delta_j=-2}
W_{\Delta_i,\Delta_j} \psi(i+\Delta_i,j+\Delta_j),
\ee
where the coefficients $W_{\Delta_i,\Delta_j}$ used for spin-1 flexion
are given in the left panel of Fig.\ts\ref{fig:grid}.  Bilinear
interpolation can be used to obtain the shear and flexion for an
arbitrary position in the field. We need to extend two grid rows and
columns on each boundary of the entire field, which means we use a
$(N+4)^2$ grid to perform the method for inner $N^2$ grid.  Note that
the flexion drops quickly with increasing distance from the cluster
center, so that there is few images with flexion signal near the grid
boundary.

Since shear and flexion are second- and third-order derivatives of $\psi$,
the potential field is not fixed with shear and flexion constraints
alone. However, $\kappa$ is unchanged under the
transformation $\psi(\vc\theta) \to \psi(\vc\theta) +\psi_0 +\vc\alpha \cdot
\vc\theta$, where $\psi_0$ and $\vc\alpha$ are arbitrary constants.
We leave the constants free to simplify the numerical solution.
The mass-sheet degeneracy transformation of the potential is given by $\psi
\to \psi'= (1-\lambda) \theta^2/2 +\lambda \psi$, which does not
affect the reduced shear and reduced flexion, but affects $\kappa$ as
\be
\kappa' = (1-\lambda) + \lambda \kappa,
\elabel{massheet}
\ee
which we call the $\lambda$ transformation.
We will take advantage of this transformation for adjusting $\kappa$
later.
We note that the mass-sheet transformation is exact only for all
sources being located at the same redshift; for a redshift
distribution, the mass-sheet degeneracy is broken. However, this
breaking is not very strong, as long as no multiple images of sources
with very different redshifts are considered \citep[see][]{2005A&A...437...39B}

\section{\label{sc:test}Numerical test of field properties}
\subsection{NIS model}
\begin{figure*}
  \centerline{\scalebox{1.1}
    {\includegraphics[width=7.5cm,height=7cm]{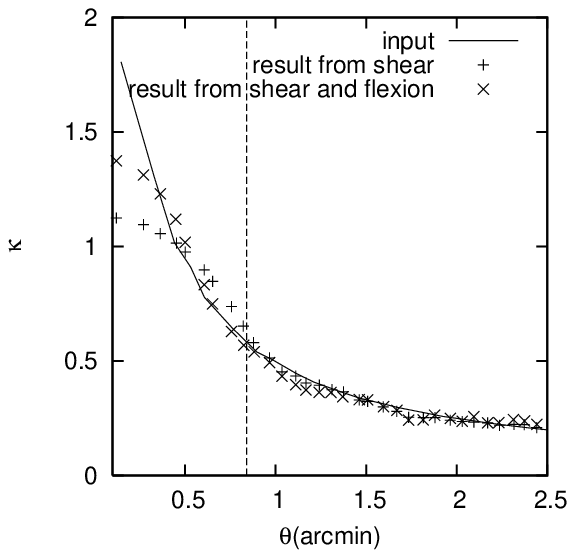}
      \includegraphics[width=7.5cm,height=7cm]{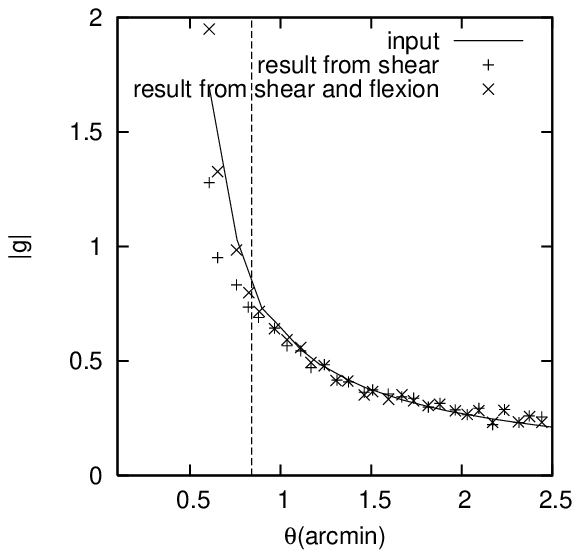}}}
  \caption{
    Radial profile of the NIS cluster, the solid line is the input model,
    the plus points are the reconstructed result with weak lensing shear
    only, and the cross points are the result with weak lensing shear and
    flexion. The mock data is generated out of the vertical line.
    Left panel: convergence $\kappa$ after $\lambda$ transformation,
    Right panel: the absolute value of reduced shear $|g|$,
    all for a source at $z_s \to \infty$.
  }
  \label{fig:kmapr}
\end{figure*}

First we perform our method on a Non-singular Isothermal Sphere model. The
data is generated on a $40\times 40$ grid potential field with
\be
\psi(\theta) = {\theta_{\rm E} \over 2} \sqrt{\theta^2 + \theta_c^2} +
{\theta_{\rm E}\theta_c \over 2} {\rm ln}
\rund{\sqrt{\theta^2+\theta_c^2} -\theta_c \over \theta},
\ee
where $\theta_{\rm E}=1$ arcmin and $\theta_c=0.2$ arcmin. The reduced
shear and reduced flexion are calculated by finite differencing.  To
ensure that no strongly lensed objects are included, we reject all the
data points of which the absolute value of the reduced shear is larger
than 0.9.  Fig.\ts\ref{fig:kmapr} shows the original and reconstructed
radial $\kappa$ and $|g|$ profiles. The value is obtained by annular
bin average on the grid field, the origianl one as well as the
reconstructed one. The small fluctuations of the results around the
input profile is mainly due to noise and low grid resolution.  In the
left panel, we plot the convergence $\kappa$ after applying the
$\lambda$ transformation Eq.\ref{eq:massheet}. We can see that
convergence is well recovered for large $\theta$, up to the vertical
line. For smaller $\theta$, the results from shear alone deviate
strongly from the input. However, using shear and flexion together
significantly improves the resulting mass profile.  In the right
panel, the absolute value of the reduced shear is shown, which is not
affected by mass-sheet degeneracy.  Both results, from shear alone and
from combining shear and flexion show agreement with the input
model. Again for small $\theta$, the result from the combined shear
and flexion data gives better agreement with the input line.
\subsection{Mock data}

Our mock data uses a cluster taken from N-body simulations by
\citet{2002ApJ...574..538J,2002MNRAS.335L..89J}. The cluster is
simulated in the framework of a $\Lambda$CDM model with
$\Omega_{\Lambda}=0.7$, $\Omega_{\rm m}=0.3$, the normalization of
power spectrum $\sigma_8=0.9$ and Hubble constant $H_0=70 {\rm km\,
  s^{-1} Mpc^{-1}}$.  Dark matter halos are identified with the
friends-of-friends method using a linking length equal to 0.2 times
the mean particle separation. The halo mass $M$ is defined as the mass
enclosed within the virial radius according to the spherical collapse
model
\citep{1996MNRAS.280..638K,1998ApJ...495...80B,2002ApJ...574..538J}.
The virial mass of the cluster which we use here is $3.4\times
10^{14}{\rm h^{-1}}M_{\odot}$ and its redshift is $z_{\rm d}=0.326$. The
particles within a box with side length of 2 virial radii were
projected onto the lens plan. The surface densities are calculated using
the smoothed particle hydrodynamics smoothing algorithm
\citep{1992ARA&A..30..543M} on a $4096\times4096$ grid. The lensing
potential is obtained using the fast Fourier transform method
\citep{1998A&A...330....1B}.

A finer grid 0.1 arcsec resolution is created around the cluster
center with side length of 4 arcmin. We obtain the mapping
$\vc\beta(\vc\theta)$
from the
lens equation for each grid point and the corresponding
second-order derivatives using cubic spline interpolation. We perform
cubic spline interpolation again to the second-order derivatives to
get the third-order derivatives.  The background sources are located
at random position in the
source planes, and their redshifts follow a Gamma
distribution
\begin{equation}
  p(z) = \frac{z^2}{2\:z_0^3} \exp\rund{-z/z_0}\; ,
\end{equation}
where $z_0 = 1 / 3$. The peak is at $z = 2 / 3$ and the mean redshift
is $\langle z \rangle = 3 z_0 = 1$. The Newton-Raphson method is used
to find the corresponding images position on the image plane.  The
reduced shear and flexion for each image were linear interpolated
using four nearest grid points.

\begin{figure*}
  \centerline{\scalebox{1.0}
    { \includegraphics[width=8.0cm,height=8.0cm]{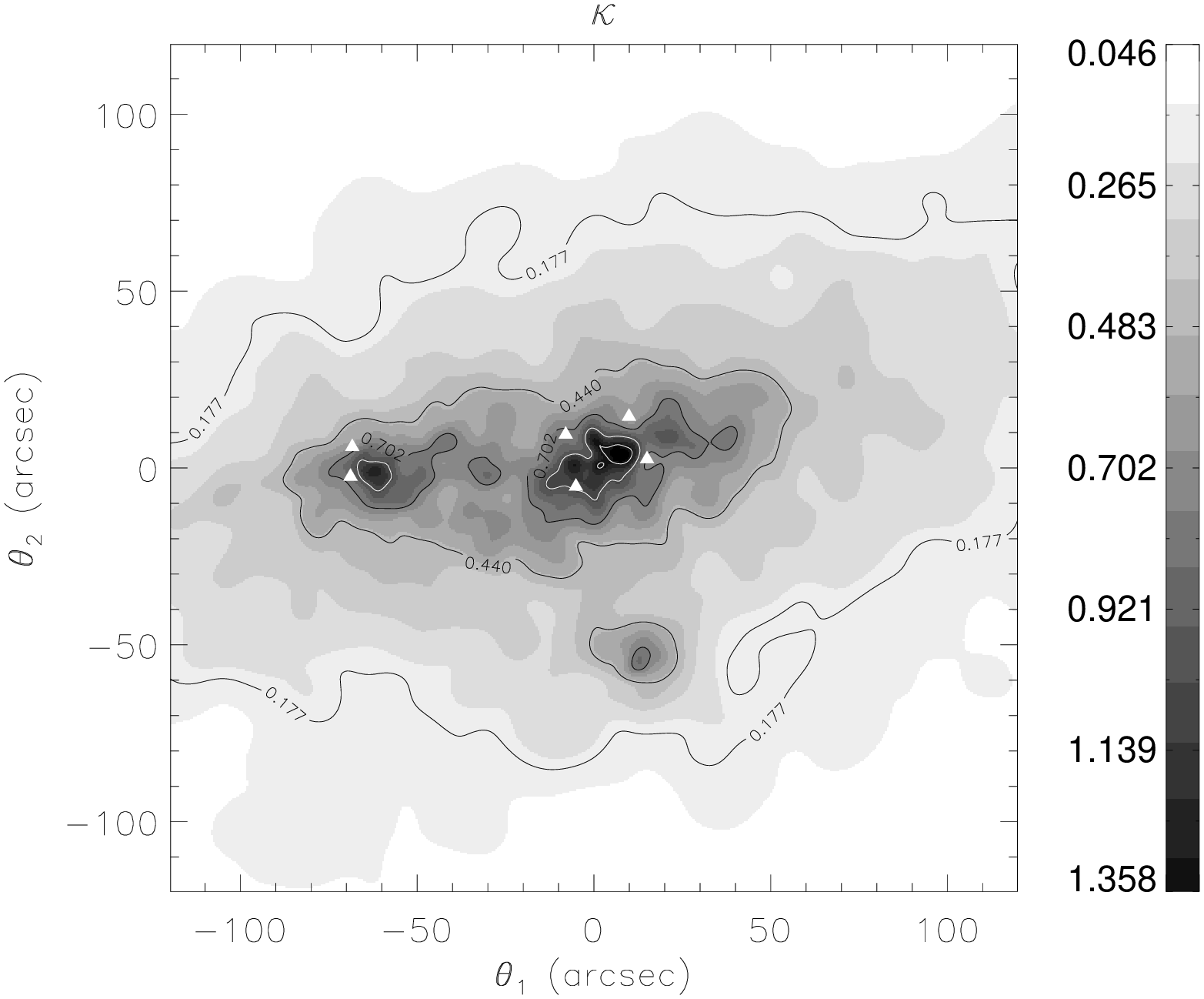}
      \hspace{1.0cm}
      \includegraphics[width=8.0cm,height=8.0cm]{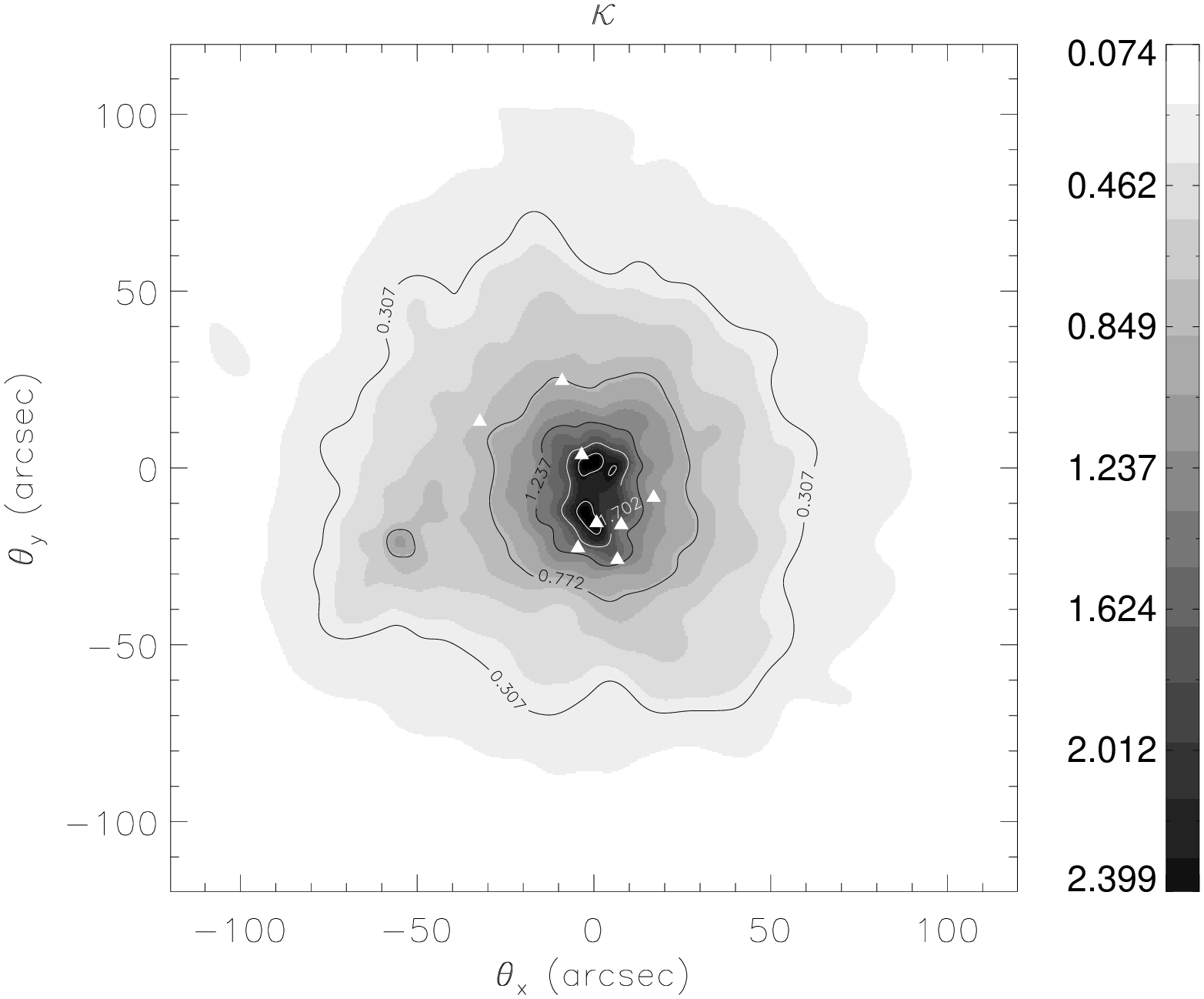}}}
  \caption{
    The convergence map of two simulated cluster used for generating mock
    strong lensing multiple images, weak lensing shear and flexion data.
    The triangles denote the image positions of multiple image
    systems which we use for the reconstruction.
    $\kappa$ is plotted for sources at $z_s\to \infty$, given in linear
    gray-scale and contours.
    We name the left panel cluster d02 and the right one cluster d03.
  }
  \label{fig:inputmap}
\end{figure*}

One simulated cluster with two different projection directions is used
to generate the two different sets of mock data
(Fig.\ts\ref{fig:inputmap}). We name the two sets of data as d02 and
d03.  All of the reduced shear data is used in our calculation, but
for flexion, we only consider reduced flexion with absolute value in
the range from 0.01 to 0.5.  For higher values, merging of multiple
images most likely render any flexion measurement impossible (in
fact, there the whole concept of flexion break down -- see Schneider
\& Er 2008).  On the other hand, very small values of the reduced
flexion are exceedingly difficult to measure and would contribute very
little information due to the low signal-to-noise.
\begin{figure*}
  \centerline{\scalebox{1.0}
    {\includegraphics[width=8.0cm,height=8.0cm]{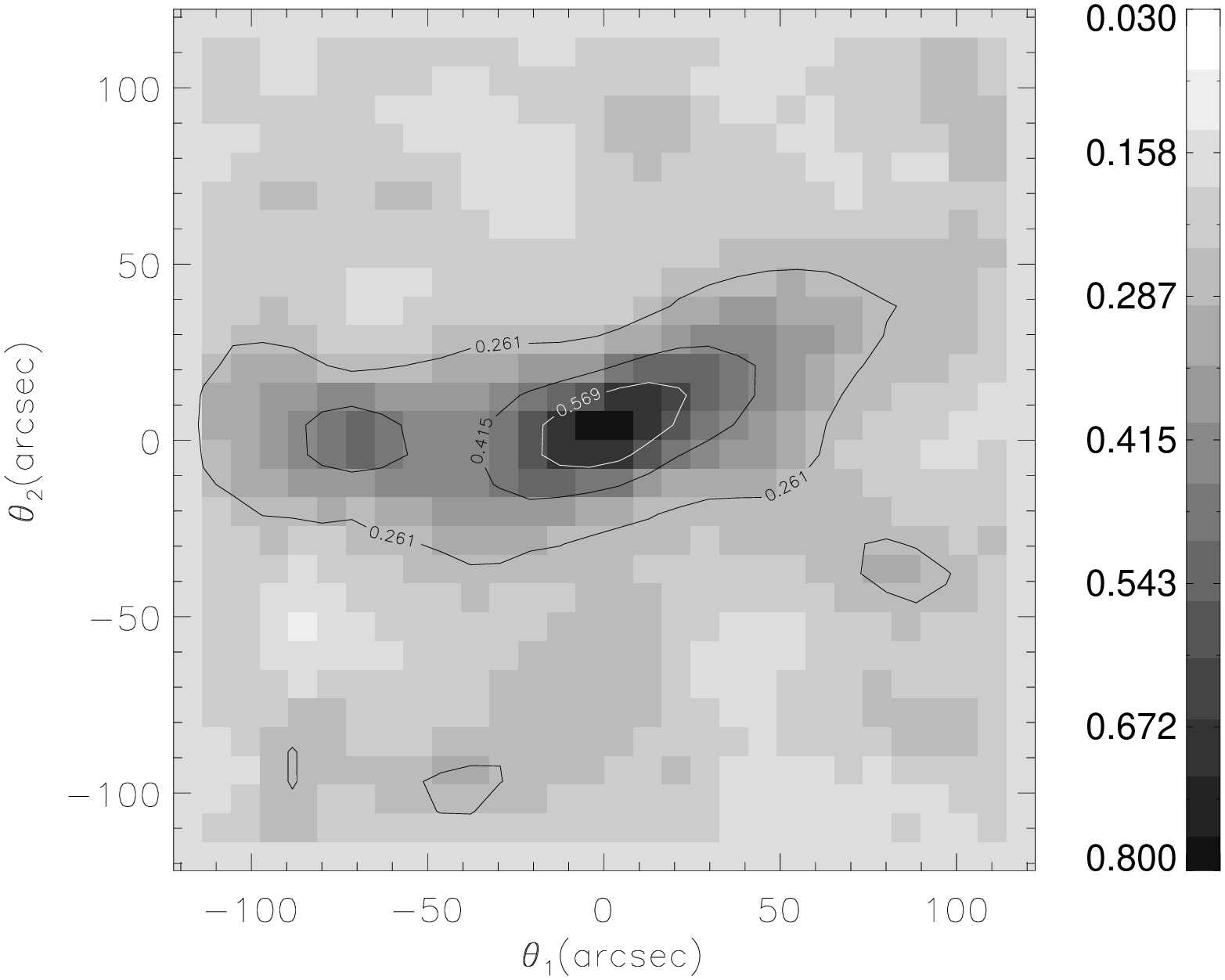}\hspace{1.0cm}
      \includegraphics[width=8.0cm,height=8.0cm]{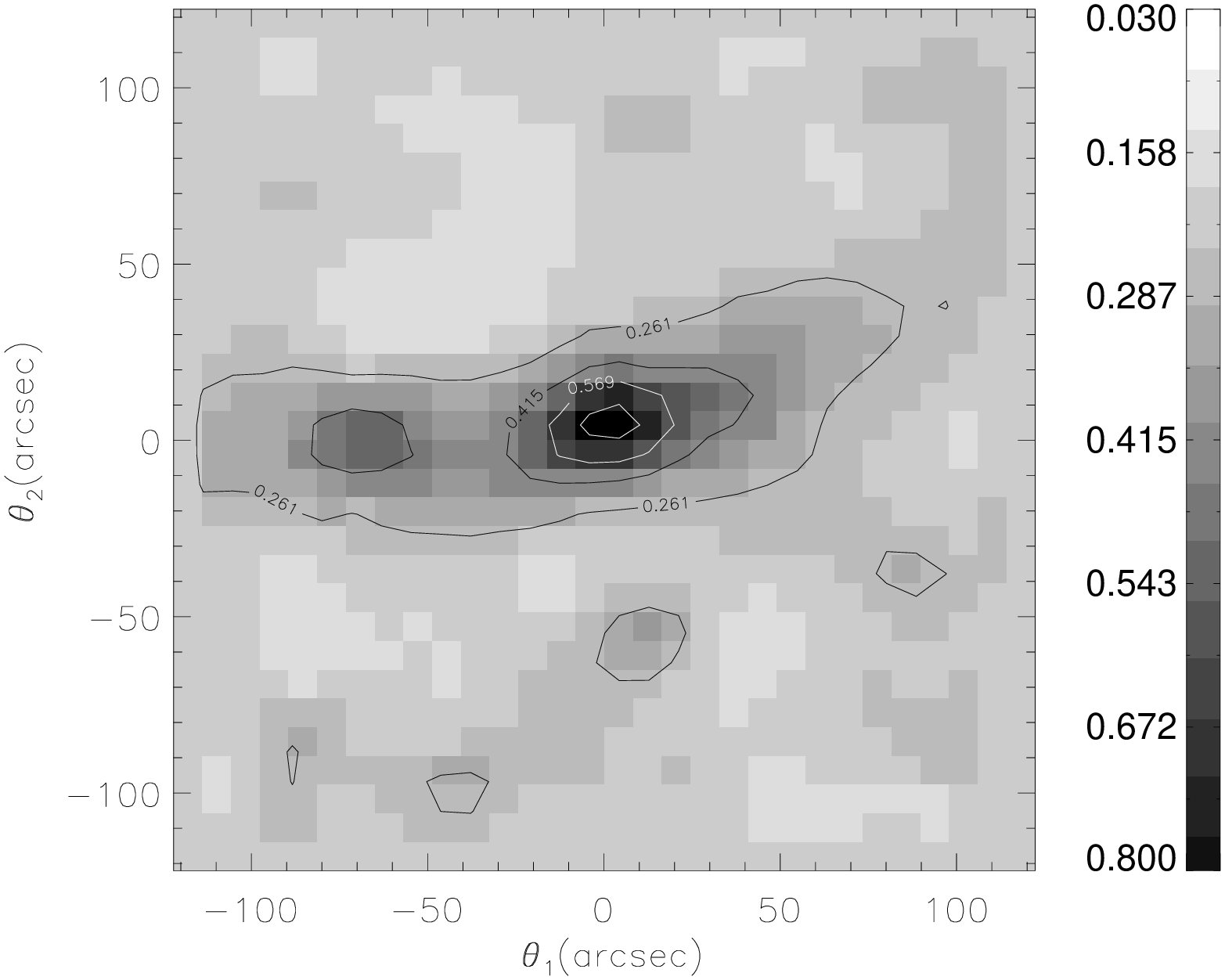}}}
  \caption{
    $\kappa$-maps obtained from shear and flexion reconstruction of the mock
    data cluster d02 after the $\lambda$ transformation
    (Eq.\ref{eq:massheet}), where $\lambda$ is chosen to yield the smallest
    $D^2$ (Eq.\ref{eq:kdif}).
    The left panel shows the result using $N_{\rm g}=1000$ shear galaxies and 3
    multiple image systems, while in the right panel we add 89 flexion
    values.
  }
  \label{fig:d02}
\end{figure*}
\begin{figure*}
  \centerline{\scalebox{1.0}
    {\includegraphics[width=8.0cm,height=8.0cm]{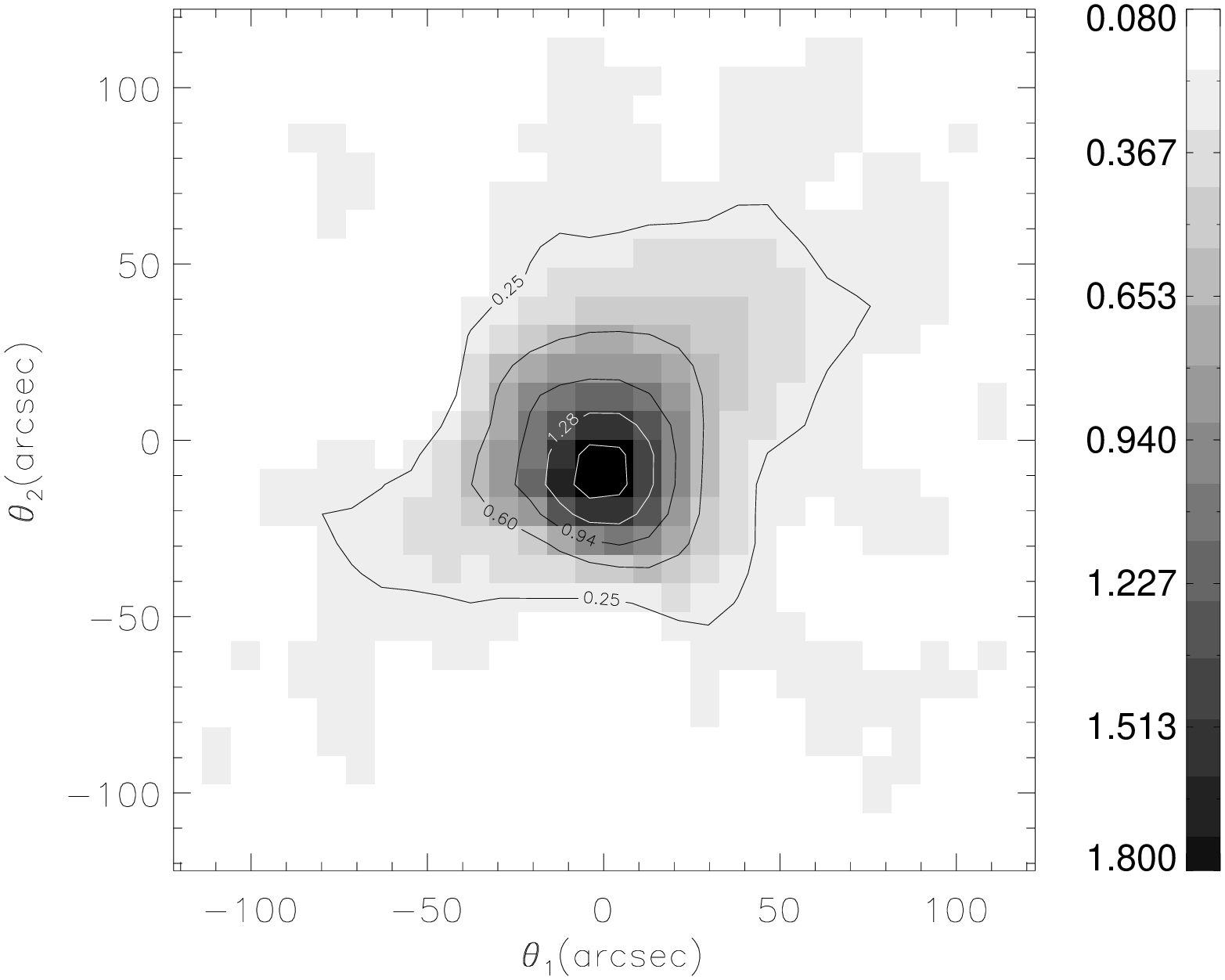}\hspace{1.0cm}
      \includegraphics[width=8.0cm,height=8.0cm]{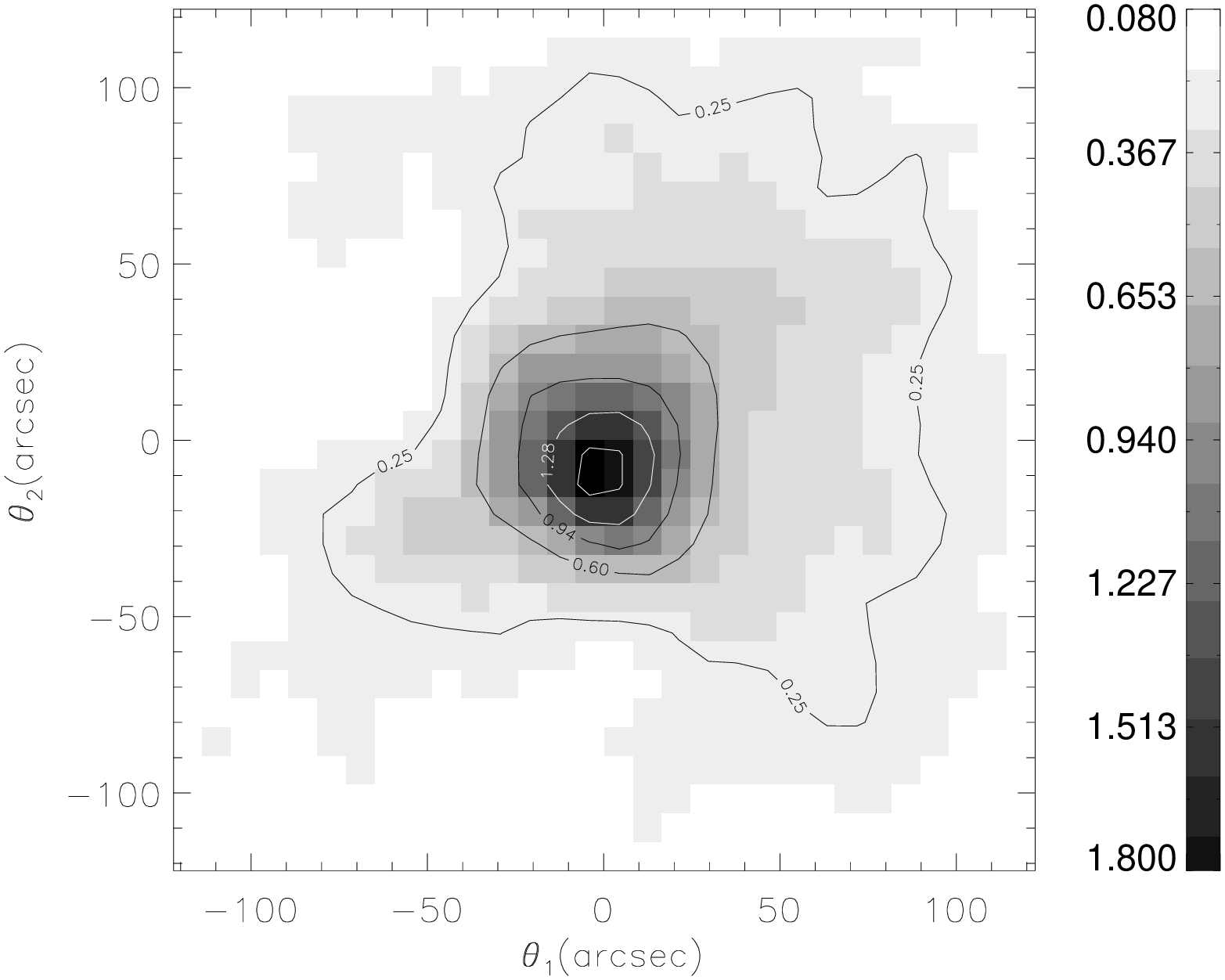}}}
  \caption{
    Reconstructed $\kappa$-maps from mock data cluster d03 after the $\lambda$
    transformation.
    Left panel shows the result using $N_{\rm g}=$ 1000 shear
    galaxies, and 3 multiple image systems, whereas the
    right panel also uses 60 flexion values.
  }
  \label{fig:d03}
\end{figure*}
\subsection{Reconstructed $\kappa$ map}
The two mock catalogues are used to test the performance of our method.
We start with an initial $20\times 20$ grid, increase $N_x$ and $N_y$ by one
for each outer loop iteration, up to a $30\times 30$ grid.
We use $N_{\rm g}=1000$ weak lensing galaxy images in each reconstruction,
which is an accessible background galaxy number density of $\sim$ 60 images
arcmin$^{-2}$. The result of the reconstructions are
shown in Fig.\ts\ref{fig:d02} for d02 and Fig.\ts\ref{fig:d03} for d03.
The initial regularization parameter is set to $\eta=200$ for cluster
d02, and
$\eta=300$ for cluster d03, and is increased by a factor of
10 for each outer-level iteration.
It is usually better to set high $\eta$ and allowing $\kappa$ to change slowly.
Since our reconstruction is done in a three-level iteration,
and in each step we ensure $\chi^2/N_{\rm dof}\sim 1$, the
method can successfully adapt to the data and the results are not sensitive to
the initial choice of $\eta$.
We need also an initial $\kappa^0$ field for the regularization.
A simple model $\kappa^0=0.01$ is used here. We have also performed
additional reconstructions with different initial models, and found the
results are nearly independent of the initial $\kappa^0$, but a more
realistic model allows for a faster convergence.

The number of flexion values that we used for the two cluster is
different. We obtained 89 reduced flexion value between 0.01 to 0.5
in d02 (Fig.\ts\ref{fig:d02}), and 60 that for d03
(Fig.\ts\ref{fig:d03}). Since there is more significant extended structure in
cluster d02 than in d03, we found that there are more high flexion signal data
(the absolute value of the spin-1 reduced flexion greater than 0.01) in
cluster d02 than that in cluster d03.

The results show that our method can reproduce the main properties of
the projected mass distribution of both clusters, and is especially
powerful in resolving the substructure.  Fig.\ts\ref{fig:d02} shows
that our method can reconstruct the $\kappa$ map by combining multiple
images, weak lensing shear and flexion data for cluster
d02. Unfortunately we cannot clearly distinguish local small-scale
maxima which are due to noise from the true low-mass substructure,
even with the help of flexion, like the one in the bottom corner of
Fig.\ts\ref{fig:d02}. However we can see that after including flexion,
the halos become more peaky and more massive substructures are
resolved and at the correct positions.  In Fig.\ts\ref{fig:d03} it is
encouraging to see that besides the overall mass distribution of the
cluster halo, our method can resolve the small clumps after including
flexion data. The small clump is not very significant even in the
input convergence (Fig.\ts\ref{fig:inputmap}), which is at a resolution
of $400\times 400$.  We also performed additional tests in which we
use different sets of weak lensing shear and flexion data for both
clusters and confirmed the validity of our method. In some cases of
cluster d03 data, the shape properties can be better reconstructed and
the small clump can be clearly resolve, but the position of the small
clump can have an offset up to 10 arcsec from its real position. In
some other realizations, the small clump cannot be clearly resolved,
which is due to the noise and local low background image density
around the small clump.

As an additional test, we calculate the difference $D^2$ between our
reconstruction and the input $\kappa$, which is defined as
\be
D^2 ={1 \over N'}\sum_{i,j}^{N'}|\kappa_{ij}-\kappa_{ij}^{\rm (input)}|^2,
\elabel{kdif}
\ee
where $N'$ is number of grid points which are not inside the critical curves.
We apply the $\lambda$ transformation (Eq.\ref{eq:massheet}) to the
convergence results, choosing $\lambda$ such as to yield the smallest difference $D^2$.
For cluster d02, the $D^2_{\rm sw}$ of the result from shear only is
0.0086, and when including flexion, it becomes 0.0085.
For cluster d03, $D^2_{\rm sw}=0.0077$ and $D^2_{\rm swf}=0.0066$,
respectively. Whereas we have employed the multiple image systems in
our reconstruction, in cases where such strong lensing systems are not
available the relative improvement obtained by the inclusion of
flexion is expected to be considerable larger.

We also enlarged the allowed range for flexion
to $[0.001,0.5]$ and find that in
some cases, $D^2$ becomes larger after including flexion.
After checking our resulting $\kappa$ on all grids, we identify the points
which give larger $D^2$ after combining the flexion signal.
We are aware that the relation of the reduced flexion and the reduced shear
(Eq.\ref{eq:reducedt1}) can transfer extra noise from shear into flexion,
especially in case of large intrinsic shear noise. In brief, if we decrease
the threshold of flexion and use more data with lower signal-to-noise. The
minimizing of $\chi^2$ will relatively lose the constrain on the region with
strong flexion signal, thus a bigger $D^2$ is obtained.


Finally a word on the dispersion of flexion $\sigma_t'$. This is a difficult
parameter to determine at the moment, since we have little knowledge about the
noise behavior in flexion measurement. The one we used in this paper
(Eq.\ref{eq:fsigma}) has a problem: as pointed out by
\citet{2009arXiv0909.5133B}, the flexion variance is biased by the content of
substructure. Moreover, we noticed that the $\sigma_{t'}$ we used is
underestimated, it can be seen from following: $\chi^2_f/N_f$ is significant
smaller than 1, i.e. $\sigma_{t'}$ is too large and the constrain from flexion
is lose and down weighted. If we apply more steps of iteration,
the noise might be over fitted. In that case, the cluster becomes very
peaky, and looks like being truncated at some edge region.
Some other forms for $\sigma_t'$ have been also tested, e.g. in
analogy to the shear
\be
\sigma^2_{t'}=\rund{1-(\theta_0|t'|)^2}^2\sigma^2_{t^s} + \sigma'^2_{\rm err},
\ee
where $\theta_0$ is size of the image. We can easily see that this
$\sigma_{t'}$ is not independent of image size and $\sigma'_{\rm err}$ is
different from that of shear (Eq.\ref{eq:esigma}), since it is not
dimensionless. It turns out that flexion is underestimated, of which
the parameter we used is $\sigma_{t^s}=0.1/''$ and $\sigma'_{\rm err}=0.1/''$.
Thus, flexion variance will affect the constraint of flexion. A more suitable
model of flexion noise need to be constructed.

\section{\label{sc:conclusion}Conclusions}
In this paper we propose a method for projected cluster mass reconstruction,
which combines strong, weak lensing shear and flexion data.
The method is based on a least-$\chi^2$ fitting of the lensing potential
$\psi$ \citep{2005A&A...437...39B,2006A&A...458..349C}.
The particular strength of this method is that the flexion data provides more
information to the inner parts of the cluster and on substructure.

We test the performance of the method on our mock clusters, comparing
the results with and without flexion. In the NIS cluster, our method
can reproduce the radial profile of the convergence and the reduced
shear. Flexion can significantly improve the result in the inner
part of the mass profile.  In the other test, we generate our mock data
from simulated clusters.  We are able to reconstruct the main
properties of the cluster mass distributions; in particular when the
flexion data is included, our method can successfully resolve the
cluster and substructure.  In addition, our result is almost
independent of the initial model $\kappa^0$ and the regularization
parameter $\eta$.

We have assumed that the intrinsic flexion is small. However, the
correction for the reduced flexion introduces extra noise from shear
to flexion, especially in the case where the intrinsic galaxies are
highly elliptical.  The effect of the noise is limited where the shear
and flexion are strong, which is, however, not the case in the outer
regions of the clusters. It is of interest
to study the relation of intrinsic noise and flexion in detail.

In \citet{2007ApJ...666...51L,2008ApJ...680....1O}, the result of mass
reconstruction by flexion has shown that flexion is sensitive to substructure,
and insensitive to the smooth component of the cluster. Our method of
combining shear and flexion takes the advantages of shear on cluster scale, and
of flexion for substructures.

The number density of background galaxies that we used, $\sim$ 60 images
arcmin$^{-2}$, is accessible by current telescopes such as HST.
With future telescopes, the accuracy and number density of images can
be improved. Once flexion can be measured accurately, the noise
behavior and intrinsic flexion scatter are certainly need to be
studied before putting flexion into practical use.

\section*{Acknowledgments}
We thank Marusa Bradac, Thomas Erben, Holger Israel, Stefan Hilbert and
Daniela Wuttke for useful discussions. We also thank Yipeng Jing for providing
the N-body simulated cluster.
XE was supported for this research through a stipend from
the International Max-Planck Research School (IMPRS) for Astronomy and
Astrophysics at the University of Bonn.GL was supported by the Humboldt
Foundation.

\begin{appendix}
\section{\label{app:a}The iteration}

We present the method outlined in Sect.\ts\ref{sc:3.1} on how we linearize and
solve Eq.(\ref{eq:chi2}).
The lensing quantities are calculated by finite differencing, and are thus
linear combinations of $\psi$ at each position. They are expressed in the
following matrix notations (Fig.\ts\ref{fig:grid})
\bea
\kappa(\theta_i) = M^{\kappa}_{ik}\psi_k;\;\\
\gamma_1(\theta_i) = M^{\gamma_1}_{ik}\psi_k;\quad
\gamma_2(\theta_i) = M^{\gamma_2}_{ik}\psi_k;\;\\
{\cal F}_1(\theta_i)=M^{\rm f1}_{ik}\psi_k;\quad
{\cal F}_2(\theta_i)=M^{\rm f2}_{ik}\psi_k;\;\\
{\cal G}_1(\theta_i)=M^{\rm g1}_{ik}\psi_k;\quad
{\cal G}_2(\theta_i)=M^{\rm g2}_{ik}\psi_k.
\eea
Then we plug these into Eq.(\ref{eq:chi2}) and obtain the 
full form of
the equations. Here we write down the first flexion term as an example.
The strong lens multiple images, shear and regularization part can be
found in \citet{2005A&A...437...39B} and the second flexion term will be
analogously to the first one,
\be
\chi_{\rm f1}^2(\psi) = \sum_{i=1}^{N_{\rm f}} {|(1-\kappa) t_1 - F|^2 \over
  (1-\kappa)^2 \sigma_{t1'}^2}\;;
\ee
here $t_1$ is given by Eq.(\ref{eq:reducedt1}), and we drop the prime on
$t_i'$ in this appendix, for notational simplicity.  We also omit the index
$i$ to all parameters of every galaxy for simplicity.  We fix the denominator
$\hat\sigma_{t1}^2 = (1-\kappa)^2\sigma_{t1}^2$ as constant at each iteration
step, so they will not appear in the derivative,
\bea {\partial \chi_{\rm f1}^2(\psi) \over \partial \psi_k} &=&
\sum_{i=1}^{N_{\rm f}}{-2 \over \hat \sigma_{t1}^2}
\sum_{r=1,2}\eck{\eck{(1-\kappa) t_{1r} - F_r } \rund{ t_{1r} {\partial \kappa
\over \partial \psi_k} + {\partial F_r \over \partial \psi_k}}} \nonumber \\
&=& \sum_{i=1}^{N_{\rm f}} {-2 \over \hat \sigma_{t1}^2} \bigg[ M^{\rm
f1}_{ij}M^{\rm f1}_{ik} + M^{\rm f2}_{ij}M^{\rm f2}_{ik}+ (t_{11}^2+t_{12}^2)
M^{\kappa}_{ij}M^{\kappa}_{ik} \nonumber \\ 
&& + t_{11}(M^{\rm f1}_{ij}M^{\kappa}_{ik} +M^{\rm f1}_{ik}M^{\kappa}_{ij})
\nonumber \\ 
&& + t_{12}(M^{\rm f2}_{ij}M^{\kappa}_{ik} +M^{\rm f2}_{ik}M^{\kappa}_{ij})
\bigg] \psi_k \nonumber \\ 
&& + {2\over \hat \sigma_{t1}^2} \eck{t_{11}M^{\rm
f1}_{ij} + t_{12}M^{\rm f2}_{ij} +(t_{11}^2+t_{12}^2) M^{\kappa}_{ij}}, \eea
where $t_{11}$ and $t_{12}$ are the two components of the spin-1 flexion
estimator of galaxy images.

It is easy to separate the terms with or without $\psi$, and write
Eq.(\ref{eq:chi2}) in the form
\be
B_{kj} \psi_k = V_j,
\ee
with the matrix $B_{kj}$ and vector $V_{j}$ containing the contributions from
the nonlinear part.
\bea
B_{kj}^{\cal F}&=&\sum_{i=1}^{N_{\rm f}}{1\over \hat\sigma^2_{t1}}
\big[ M^{\rm f1}_{ij}M^{\rm f1}_{ik} + M^{\rm f2}_{ij}M^{\rm f2}_{ik} +t_{11}
  (M^{\rm f1}_{ij}M^{\kappa}_{ik} +M^{\rm f1}_{ik}M^{\kappa}_{ij}) \nonumber\\
  &+& t_{12}(M^{\rm f2}_{ij}M^{\kappa}_{ik} +M^{\rm f2}_{ik}M^{\kappa}_{ij}) +
  (t_{11}^2+t_{12}^2) M^{\kappa}_{ij}M^{\kappa}_{ik}\big],
\eea
where $i$ denote summation over all the galaxy images. The data vector is
\be
V_j^{\cal F}=\sum_{i=1}^{N_{\rm f}} {1\over \hat\sigma^2_{t1}}
\eck{t_{11}M^{\rm f1}_{ij} + t_{12}M^{\rm f2}_{ij} +(t_{11}^2+t_{12}^2)
  M^{\kappa}_{ij}}.
\ee
Here $B_{kj}^{\cal F}$ and $V_{j}^{\cal F}$ are contributions from spin-1
flexion.  The same calculation can be performed for spin-3 flexion, and a
similar result is obtained.  Combining all the matrixes and data vectors, one
can complete the matrix $B_{kj}$ and the vector $V_j$.

\end{appendix}

\bibliographystyle{aa}
\bibliography{/users/xer/bib/refbooks,/users/xer/bib/reflens,/users/xer/bib/flexion,/users/xer/bib/refcos,/users/xer/bib/simcos}

\end{document}